\documentclass[11pt]{article}
\usepackage{amsmath,amsthm,amssymb,amsbsy,epsfig,fancyhdr,calc,ifthen,float,psfrag,mathrsfs,graphics,color,fullpage}
\restylefloat{figure}
\usepackage{pdfpages}
\usepackage[noblocks]{authblk}
\usepackage[font=small,labelfont=bf]{caption}
\usepackage{enumitem}

\title{Universal features in panarthropod \\ inter-limb coordination during forward walking}% Force line breaks with \\

\author[1,2]{\normalsize Jasmine A. Nirody}
\affil[1]{\small Center for Studies in Physics and Biology, Rockefeller University, New York, NY 10065 USA}
\affil[2]{\small All Souls College, University of Oxford, Oxford OX1 4AL United Kingdom}

\date{\today}% It is always \today, today,
\begin{document}
\maketitle
\begin{abstract}
Terrestrial animals must often negotiate heterogeneous, varying environments. Accordingly, their locomotive strategies must adapt to a wide range of terrain, as well as to a range of speeds in order to accomplish different behavioral goals. Studies in \textit{Drosophila} have found that inter-leg coordination patterns (ICPs) vary smoothly with walking speed, rather than switching between distinct gaits as in vertebrates (e.g., horses transitioning between trotting and galloping). Such a continuum of stepping patterns implies that \textcolor{black}{separate neural controllers are not necessary} for each observed \textcolor{black}{ICP}. Furthermore, the spectrum of \textit{Drosophila} stepping patterns includes \textcolor{black}{all canonical coordination patterns observed during forward walking in insects}. This raises the exciting possibility that the controller in \textit{Drosophila} is common to all insects, and \textcolor{black}{perhaps more generally to panarthropod walkers}. Here, we survey and collate data on leg kinematics and inter-leg coordination relationships during forward walking in a range of arthropod species, as well as include data from a recent behavioral investigation into the tardigrade \textit{Hypsibius exemplaris}. Using this comparative dataset, we point to several functional and morphological features that are shared amongst panarthropods. The goal of the framework presented in this review is to emphasize the importance of comparative \textcolor{black}{functional} and morphological analyses in understanding the origins and diversification of walking in Panarthropoda.
\end{abstract}

\begin{figure}
\includegraphics[width=\textwidth]{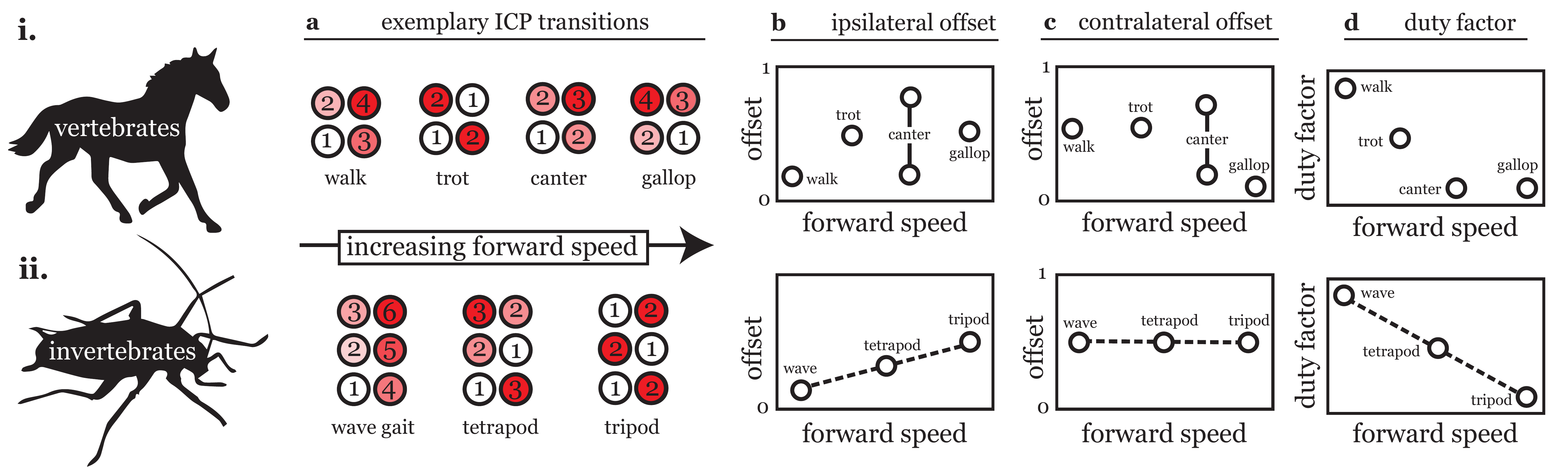}
\caption{\textit{Transitions in stepping pattern with walking speed in vertebrates \textbf{(i)} and invertebrates \textbf{(ii)}.} Both vertebrates and invertebrates show changes in inter-leg coordination patterns (ICPs) with speed.  \textcolor{black}{While our discussion focuses on leg kinematics, it is important to note that gait transitions comprise both changes in leg coordination and body dynamics, as has been extensively documented in vertebrate species \cite{alexander1989optimization}. Representative ICP transitions for (i) tetrapods and (ii) hexapods are shown.}  \textbf{(a)} Horses transition from a walk at low speeds, to a trot at intermediate speeds, to a canter or gallop at high speeds, while insects switch from pentapodal wave to tetrapodal and then tripod coordination as walking speed increases. Numbering denotes the order of footfalls within a full stride cycle; timing of footfalls is also denoted from lighter to darker coloring. Stepping patterns in vertebrates can be categorized into discrete gaits that \textcolor{black}{are mirrored by transitions in body dynamics driven by energetic considerations.} These transitions show discontinuities in parameters such as phase offset between leg pairs \textbf{(b, c)} and duty factor \textbf{(d)}. Note that, as a three-beat gait, canter is asymmetric and exhibits different characteristic phase offset between the two ipsilateral and contralateral leg pairs.  In contrast, invertebrate walking is a continuum with intermediate stepping patterns providing smooth transitions between `canonical' stepping patterns. \textcolor{black}{Though the `canonical' stepping patterns shown here correspond to hexapods, observed trends for phase offsets and duty factor are generalizable to arthropods with any number of legs:  \textbf{(b)} ipsilateral phase offset increases continuously with walking speed, \textbf{(c)} contralateral phase offset is anti-phase across speeds, and \textbf{(d)} duty factor decreases continuously with walking speed \cite{manton1952evolution}.} }
\label{fig:vertinvertcomp}
\end{figure}

Walking, a behavior fundamental to numerous tasks important for an organism's survival, is assumed to have become highly optimized during evolution. Terrestrial animals must navigate rough, varying landscapes; as such, stepping patterns must be flexible in order to successfully complete a range of behavioral goals across a range of terrains. The foremost of these adaptations is variability in the temporal and spatial coordination between leg movements. In \textcolor{black}{vertebrates}, this variability manifests as distinct gaits: for instance, a horse will switch from a walk to a trot to a gallop as it increases forward speed (Figure \ref{fig:vertinvertcomp}). These switches are generally \textcolor{black}{driven by energetic considerations} and accompanied by \textcolor{black}{changes in the movements of the animal's center of mass} as well as discontinuities or sharp transitions in at least one parameter, e.g., duty factor or the phase offset between leg pairs \cite{alexander1983dynamic,alexander1989optimization}. 

%Such gait transitions have been explicitly characterized in vertebrates \cite{alexander1983dynamic,alexander1989optimization}, but the existence of discrete gaits in legged invertebrates remains a subject of some debate in the literature.

At first glance, similar transitions with walking speeds \textcolor{black}{are} present in arthropod species. \textcolor{black}{Slow walking insects largely use a wave coordination, in which at most one leg is lifted (in `swing' phase) at a time. Insects walking at intermediate speeds utilize tetrapodal stepping patterns, in which two limbs enter swing phase simultaneously. Finally, fast-running insects employ tripod coordination, in which two pairs of three legs each lift off in sequence; each tripod comprises an ipsilateral front and hind leg and the contralateral middle leg. A schematic illustrating these canonical patterns is shown in Figure~\ref{fig:vertinvertcomp}. While inter-leg coordination patterns (ICPs) in insects are often referred to as `gaits' in the literature \cite{nishii2000legged,durr2004behaviour,bender2011kinematic}, it has yet to be explicitly shown that transitions between invertebrate ICPs with speed constitute transitions between discrete gaits \cite{alexander1989optimization}.}

The vast majority \textcolor{black}{of recent studies on arthropod locomotion consist of deep investigation into the behavior of a single organism} (most commonly, an insect). Within this framework, our understanding of the nature of \textit{transitions} \textcolor{black}{between invertebrate ICPs} is hindered by the fact that \textcolor{black}{most species utilize a limited range of spontaneous walking speeds under constrained laboratory conditions (e.g., forward walking on flat, uniform terrain). Such controlled trials do not allow for the observation of switches between preferred stepping patterns.} Ants, for instance, \textcolor{black}{have been recorded using primarily} tripod \textcolor{black}{coordination} across a speed range of approximately 5 - 30 body lengths/s \cite{reinhardt2014level,wahl2015walking,pfeffer2019high}; little data is available at lower speeds that may call for different preferred stepping patterns. Adult stick insects, on the other hand, scarcely walk on flat surfaces at speeds above 1 body length/s and thus rarely \textcolor{black}{have been observed displaying} tripodal coordination patterns \textcolor{black}{in the laboratory} \cite{cruse1990mechanisms,graham1972behavioural}.

\textcolor{black}{Accordingly}, this framework has led to the development of several models of walking, each derived from the behavior of a single, highly specialized organismal system \cite{ayali2015comparative}. For example, behavioral studies conducted in slow-walking stick insects have suggested that a small set of local coordination rules suffices to explain observed ICPs~\cite{cruse1990mechanisms,durr2004behaviour}. Inter-segmental neural pathways have also been shown to be important in coordinating leg movements in fast tripodal walkers like the cockroach \textit{Periplanta americana} \cite{pearson1973nervous}, but a clear connection between these postulated mechanisms has not been rigorously characterized. 

Studies on species that exhibit a wide range of preferred walking speeds \textcolor{black}{in the laboratory} have been useful in connecting the mechanisms underlying slow and fast walking. One such organism is the fruit fly \textit{Drosophila}, a species for which tool availability is an added benefit: \textit{Drosophila}'s status as a model organism allows for the collection of large datasets and tractable genetic manipulation of neural signals \cite{mendes2013quantification,wosnitza2013inter,szczecinski2018static}. \textcolor{black}{These studies have shown that \textit{Drosophila} show inter-leg coordination patterns that fall along a speed-dependent continuum containing all the `canonical' stepping patterns observed in other insects \cite{deangelis2019manifold}.}

Excitingly, these findings have strong implications for our understanding of the underlying locomotor control circuits and corroborate theoretical investigations suggesting that the same circuit may be able to generate the entire observed range of ICPs in \textit{Drosophila} (that is, there are not separate dedicated controllers for, e.g., tripod coordination)~\cite{wosnitza2013inter,schilling2020decentralized}. Furthermore, the stepping patterns characterized in slow- and fast-walking \textit{Drosophila} closely matched those in both stick insects and cockroaches, respectively \cite{deangelis2019manifold,schilling2020decentralized}. Importantly, this leads to the hypothesis that the underlying \textcolor{black}{control circuit} responsible for generating the spectrum of ICPs observed in \textit{Drosophila} may be common to all insects, and \textcolor{black}{perhaps more generally to all panarthropods. This hypothesis is consistent with early observations that stepping patterns in Onychophora (velvet worms, which along with Tardigrada and Arthropoda, comprise Panarthropoda) are `sufficiently wide to provide a common origin for all the more specialized types of arthropodan gait' \cite{manton1952evolution}.}

A simple model put forward based on behavioral analyses in \textit{Drosophila} suggests that walking involves connections between the neuropil of the ventral nerve cord (VNC) \cite{deangelis2019manifold}. \textcolor{black}{The arthropod central nervous system (CNS) shares a common blueprint, consisting of a brain and a series of segmented bilateral ganglia from which lateral nerves extend into each body segment and appendages \cite{niven2008diversity,yang2016fuxianhuiid,smarandache2016arthropod}. This topology is largely conserved throughout Arthropoda, though it is important to note that there exists significant diversity in ganglionic structure among arthropod classes within this general framework. For instance, crustacean ganglia are not completely fused at the midline and display a ladder-like structure, in which hemiganglia are connected by axons within each segment~\cite{storch2009crustacea}.} To this end, integrative studies that consider both functional and phylogenetic relationships among various organismal systems are vital to our understanding of invertebrate walking~\cite{ayali2015comparative}.

\textcolor{black}{In this review, we gather kinematic data on arthropod forward walking on flat surfaces. We note that our analysis is limited by data availability in the literature in two ways. First, the majority of our discussion emphasizes walking kinematics in insects, which is simply a reflection of the distribution of past research in the field; in particular, recent work emphasizing the collection of large kinematic and behavioral datasets has focused almost exclusively on insect species. We attempt to include examples (and data) from a diversity of non-insect arthropods whenever possible. We present these alongside results from our investigations in the eutardigrade \textit{Hypsibius exemplaris}~\cite{nirody2021inter}. We root our comparisons in a review of nervous system diversification across Panarthropoda \cite{niven2008diversity,smarandache2016arthropod}, noting in particular the similarities in VNC topology between tardigrades and arthopods \cite{yang2016fuxianhuiid}. We further describe several exceptions (e.g., `galloping' in some beetles \cite{smolka2013new}) that diverge from the `canonical' patterns as systems of interest for developing insight into possible adaptive mechanisms for performance in challenging environments.} 

\textcolor{black}{Second, we constrain our discussion to inter-leg coordination patterns, rather than `gaits'; it is important to note that true gaits cannot be defined simply by leg kinematics but must also take the animal's inertia into account. Gait transitions are driven by energetic considerations and must be accompanied by changes in body dynamics \cite{hoyt1981gait}. Recent studies suggest that transition between ICPs in invertebrates may similarly be driven by an optimization against physical constraints \cite{nishii2000legged,szczecinski2018static}. However, data concerning changes in center of mass (COM) dynamics in the literature are available only for a limited number of arthropod species (see, e.g., \cite{ting1994dynamic,full1990mechanics,full1991mechanics,dallmann2017load}). Given this, our analysis centers on inter-limb coordination, for which large datasets are more readily available \cite{wosnitza2013inter,mendes2013quantification,deangelis2019manifold}. This focus is encouraged by recent work suggesting that, in addition to mechanical considerations, animals with small circuits for controlling limbs may prefer particular stepping patterns that rely on simple underlying control \cite{deangelis2019manifold}.} With this work, we hope to highlight the value of performing comparative functional and morphological studies -- and, accordingly, the importance of making organismal data open and accessible -- in illuminating the origins and evolution of invertebrate walking patterns.

\section*{Methods}

Data for arthropod species in Figures \ref{fig:dutyfactor} and \ref{fig:coord} were extracted from published articles as cited. For some articles, tabular data was not available; in these cases, data was extracted from paper figures using the R package \textit{digitize}~\cite{digitize}. For inter-leg phase offsets, only mean values are shown for all species other than \textit{Drosophila}, due to the large variation in data availability across studies. 

For studies in which distributions of phase offsets between leg pairings were reported, distributions were tested against the normal distribution using the Kolmogorov-Smirnov test at the 5\% significance level. Mean values for ipsilateral phase relationships are reported only if phase offsets were normally distributed. This is due to the fact that a joint distribution of inter-leg phase offset and walking speed was rarely available, and so we attempt to avoid averaging offsets over a large range of walking speeds (e.g., pooling data from tripod and tetrapod coordination patterns). All available contralateral phase distributions showed a single peak and were normally distributed. 

All data shown for the tardigrade \textit{H. exemplaris} was collected as reported in our previous work~\textcolor{black}{\cite{nirody2021inter}}. This data will be made available at http://www.github.com/jnirody/waterbears; all other digitized data shown will be available at http://www.github.com/jnirody/invertICPs. 

\section*{Results and Discussion}

\subsection*{Invertebrate kinematics vary smoothly with walking speed}

\begin{figure}[t!]
\begin{center}
\includegraphics[width=\textwidth]{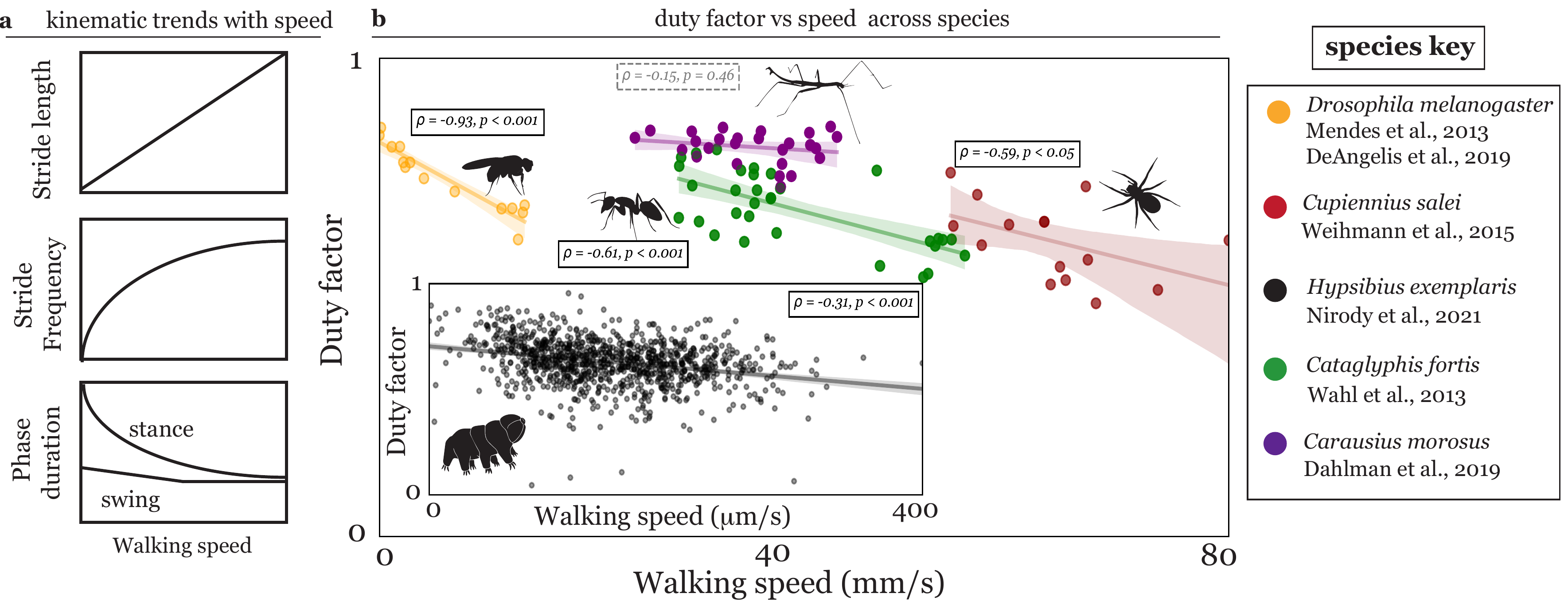}
\caption{\textit{Relationship between kinematic parameters and forward walking speed across panarthropods.} \textbf{(a)} Schematic depicting generalized relationships between walking speed and stride length (top), stride frequency (middle), and relative stance/swing duration (bottom). Both stride length and frequency increase with walking speed, \textcolor{black}{with stride frequency plateauing at high speeds \cite{manton1952evolution}}. Each step is composed of a swing (leg lifted) and stance (leg on the ground) period. \textcolor{black}{Speed is largely modulated by changes in stance duration; in contrast, swing duration remains relatively constant over walking speeds, decreasing slowly with walking speed at low to medium speeds and leveling off at high speeds . Recent work has shown little correlation between swing duration and walking speed \cite{wosnitza2013inter,nirody2021inter}} \textbf{(b)} The relative modulation of swing and stance duration within a stride is characterized  by the duty factor, or proportion of a stride spent in stance. All organisms surveyed display a smoothly decreasing duty factor with forward walking speed. Linear regression fits are shown as solid lines alongside 95\% confidence intervals; fits for which $p > 0.05$ are shown as dotted lines.}
\label{fig:dutyfactor}
\end{center}
\end{figure}

Organisms walk in order to complete a variety of behavioral goals, and must be able to do so successfully in a variety of natural environments. To this end, virtually all legged animals have developed strategies to modulate several performance metrics, including, importantly, walking speed \cite{heglund1974scaling,heglund1988speed}. While some kinematic trends with walking speed are generalizable across invertebrate species (Figure \ref{fig:dutyfactor}a), there are several distinct differences in how different species utilize the interplay between the tuning of various temporal and spatial parameters. 

Two intuitive candidates for such parameters are \textit{stride length} and \textit{stride frequency}. Like quadrupeds and bipeds (including humans), invertebrates tune both the length of their steps and the amount of time devoted to each step in order to modulate their speed of locomotion. \textcolor{black}{Stride length generally shows a linear relationship with speed across walking speeds \cite{wosnitza2013inter,weihmann2017speed,szczecinski2018static,clifton2020uneven}.} The maximum stride length achievable by an organism is dictated by absolute leg length, unless stride length can be further increased by inserting aerial phases into the \textcolor{black}{stepping pattern}. While aerial phases are commonly observed in vertebrate species (e.g., in horse trots or human running), fast-running insects almost always maintain a grounded alternating tripod pattern over a wide range of speeds \cite{full1991mechanics,goldman2006dynamics,wosnitza2013inter,reinhardt2014level,wahl2015walking,pfeffer2019high,dallmann2019motor}. Only rare instances of aerial phases in high-speed running have been observed in certain individuals (cockroach, \textit{Periplanta americana}: \cite{full1991mechanics}; ant, \textit{Cataglyphis fortis} \cite{wahl2015walking}; spiders, \textit{Hololena adnexa} and \textit{Hololena curta} \cite{spagna2011gait}). \textcolor{black}{Arthropods with higher leg numbers (e.g., arachnids, myriapods) can reach even greater speeds than hexapods during grounded running~\cite{spagna2011gait,manton1952chilopoda}.}

To increase stride frequency, organisms can reduce step cycle period by either shortening the \textit{swing} or \textit{stance} phase of the cycle. Each leg's stride comprises a protraction (swing), in which the leg is lifted and takes a step, and a retraction (stance), in which the leg is in contact with the ground and generates propulsion. \textcolor{black}{Walking speeds across panarthropod species are mainly modulated by stance duration. In contrast, swing duration generally decreases only slightly with speed at low to medium speeds and is constant at high speeds (Figure \ref{fig:dutyfactor}a; see also, e.g., \cite{wosnitza2013inter,mendes2013quantification,durr2018motor}).} This observed trend has leant support to the idea that mechanically mediated load-based coordination is a widespread control strategy \cite{szczecinski2018static}.  

This relative modulation is cleanly characterized by changes in the \textit{duty factor} -- the proportion of a cycle spent in stance phase. Transitions between discrete ICPs are often characterized by sudden changes in the duty factor: for instance, the walk-trot transition in horses is accompanied by a sharp drop in the animal's duty ratio from approximately 0.6 to 0.5 \cite{hoyt2006relations,starke2009walk}. In line with the hypothesis that insect walking lies along a speed-dependent continuum, all panarthropods surveyed (including several insect species, crustaceans, spiders, and tardigrades) during forward walking on flat surfaces show a smooth, continuous relationship between duty factor and walking speed (Figure \ref{fig:dutyfactor}b). \textcolor{black}{Arthropods with a large number of legs display significantly lower duty factors than hexapods can achieve at the highest walking speeds; for instance, some species of myriapods have been observed to run with only three out of forty legs in stance phase, corresponding to a duty factor of 0.075 (data not shown; see \cite{manton1952chilopoda,manton1954diplopoda,manton1972hexapod} for further details).}

\subsection*{Swing-stance relationships generate smooth transitions}
\begin{figure}[t!]
\begin{center}
\includegraphics[width=\textwidth]{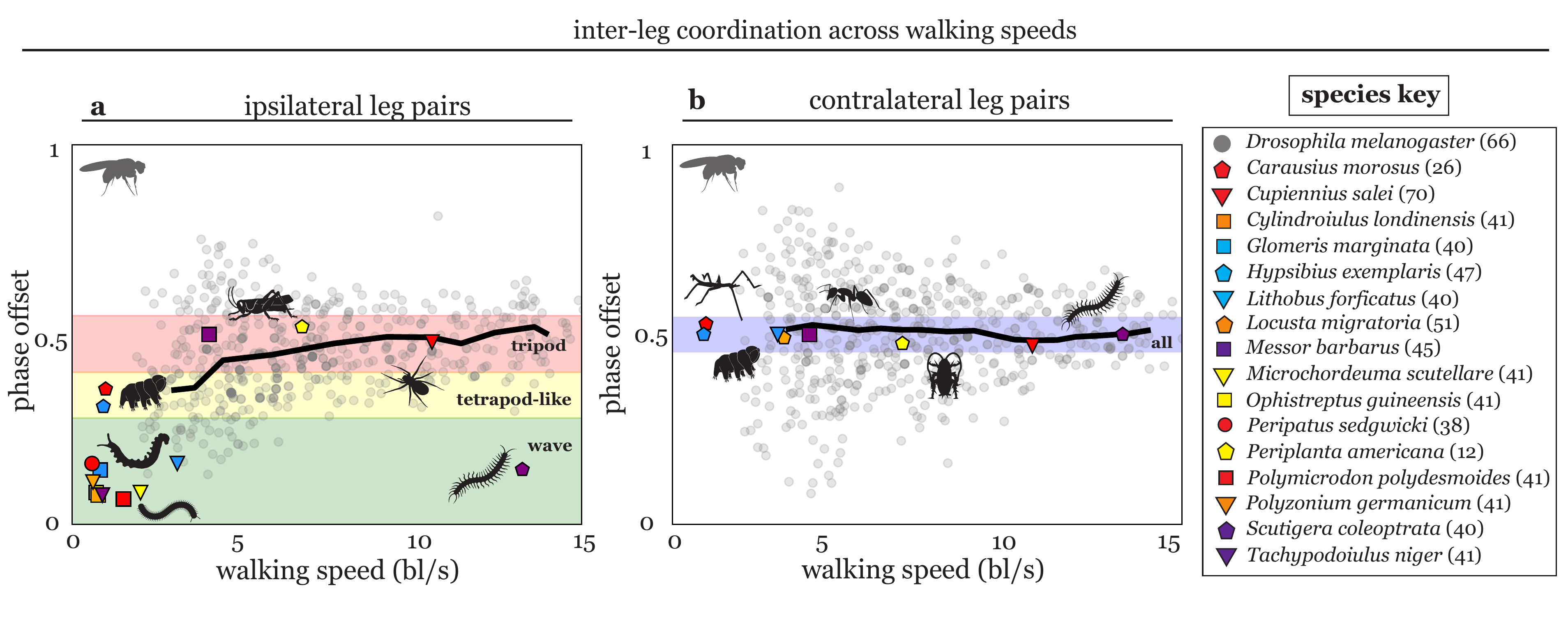}
\caption{\textit{Changes in inter-leg coordination with walking speed.} Relationship between walking speed and measured phase offset in swing initiations between \textbf{(a)} ipsilateral and \textbf{(b)} contralateral leg pairs. Ipsilateral phase relationships are reported with respect to a posterior reference leg and anterior observed leg (e.g., reference leg L3, observed leg L2). Full distribution is reported for \textit{Drosophila} (grey dots); data from \cite{szczecinski2018static}; running mean for \textit{Drosophila} is shown as a solid black line. Mean values are reported in other species; data from papers as cited in the key. For studies in which distributions were made available, only means from normally-distributed phase offsets are reported (see Methods). \textcolor{black}{Shaded regions show expected phase offsets for characterized ICPs; note that labels correspond to ICPs named in hexapodal locomotion. Wave coordination shows $\phi_I = \frac{1}{6}$ in hexapods; more generally, metachronal waves in animals with $n$ legs can show far lower phase offsets, up to a lower limit of $\phi_I \geq \frac{1}{n}$. Ipsilateral offsets close to 0 are observed for instance in slow moving millipedes \cite{manton1954diplopoda}. Tetrapod-like coordination in hexapods shows characteristic phase offsets of $\phi_I = \frac{2}{6} = \frac{1}{3}$. At fast speeds, many arthropods utilize a stepping pattern in which two consecutive sets of legs lift off in sequence ($\phi_I = \frac{1}{2}$) ; this corresponds to, for instance, tripod coordination in hexapods ($\phi_I = \frac{3}{6}$). Fast-running maxillopeds, including several species of centipedes, are able to utilize wave coordination to achieve speeds far higher than possible in hexpods. Note that all characterized patterns across speeds and body plans maintain anti-phase contralateral coupling, $\phi_C = \frac{1}{2}$.}}
\label{fig:coord}
\end{center}
\end{figure}

Changes in locomotor output are not limited to tuning the movements of single legs, but also include shifts in the temporal coordination \textcolor{black}{among} legs. Inter-leg coordination parameters are thought to be of secondary importance with respect to modulation of walking speed, but are essential for static and dynamic stability \cite{szczecinski2018static}. Though the literature often refers to `gaits' in insects, there is little evidence that invertebrates show discontinuous transitions in kinematics across forward walking speeds (Figure \ref{fig:dutyfactor}). \textcolor{black}{As noted previously, proper characterization of gaits requires consideration of the animal's inertia in addition to stepping pattern; we have considered only the latter here. As such, we limit our current discussion to ICPs only; understanding invertebrate `gaits' (whether or not they exist) will require greater availability of public data on COM dynamics for a range of arthropod species.}
\begin{figure}[t!]
\begin{center}
\includegraphics[width=0.65\textwidth]{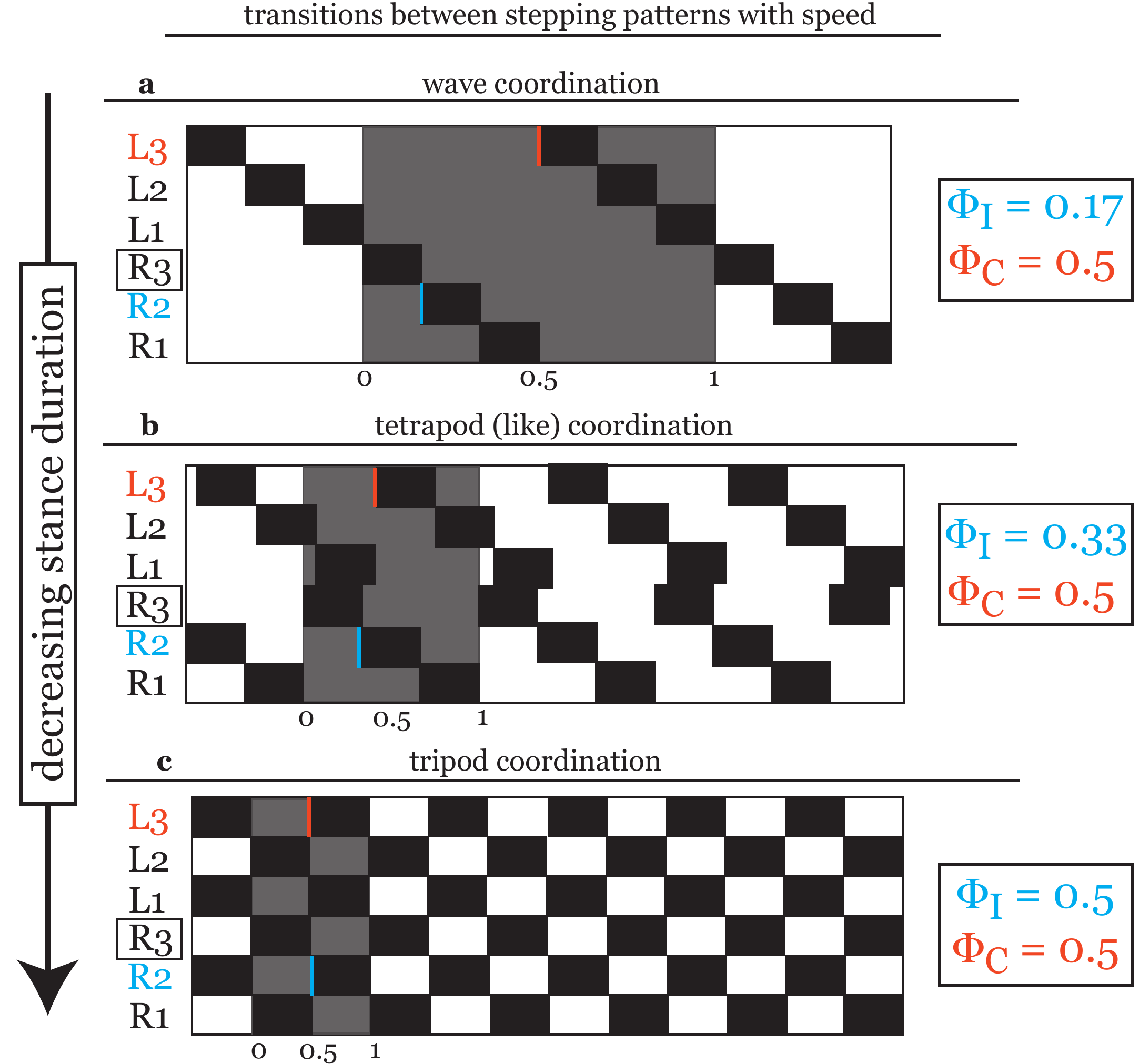}
\caption{\textcolor{black}{\textit{Observed hexapod stepping patterns form a speed-dependent continuum of ICPs generated by modulating a single parameter, stance duration.} During forward walking, arthropods transition through a spectrum of ICPs with walking speeds by modulating a single parameter: the duration of stance (duration of ground contact time) \cite{deangelis2019manifold}. Each ICP is defined by a characteristic set of phase offsets between ipsilateral ($\phi_I$, blue) and contralateral ($\phi_C$, orange) leg pairs. Ipsilateral phase offsets increase with forward walking speed, saturating in most arthropod species at $\phi_I = 0.5$; contralateral phase offset $\phi_C = 0.5$ remains constant across walking speeds. In hexapods, three `canonical' stepping patterns along this spectrum have been characterized: \textbf{(a)} wave coordination at slow speeds to  \textbf{(b)} tetrapodal coordination at intermediate speeds to \textbf{(c)} tripod coordination at high speeds. Footfall diagrams show temporal sequence of ground contacts for these three observed patterns. A full cycle [0,1] with respect to reference leg R3 is shown highlighted in grey. Swing (duration of a cycle for which a leg is lifted) is shown in black; stance  is shown in white. The relative phase offset of swing initiations by the ipsilateral anterior leg (R2, blue) and the contralateral leg (L3, orange) are denoted for each ICP. Note that a `canonical' tetrapod pattern comprises a sequence of simultaneous lift-offs by three sets of two legs. This results in a contralateral offset of $\phi_C = \frac{1}{3}$ (or $\phi_C = \frac{2}{3}$ for the mirror-image tetrapod). However, a cross-body offset in step timing such that limbs that are meant to swing simultaneously are actually slightly offset in time results in a tetrapod-like stepping pattern that shows the anti-phase contralateral phase relationship consistent with the observed continuum \cite{deangelis2019manifold}.}}
\label{fig:transitions}
\end{center}
\end{figure}
Studies in walking \textit{Drosophila} show inter-leg coordination patterns that merge together into a speed-dependent continuum. Slow-walking flies move with a pentapodal wave \textcolor{black}{coordination}, in which only one leg is in swing phase (lifted off the ground) at a time. At higher speeds, flies adopt a tetrapodal stepping pattern, in which two legs are in swing simultaneously. At the fastest speeds, flies almost exclusively utilize tripod coordination, in which two pairs of three legs swing in sequence (Figure \ref{fig:transitions}). The large variation observed in \textit{Drosophila} ICPs precludes the existence of sharp switches in coordination at characteristic speeds (Figure \ref{fig:coord}); instead, flies often can make use of multiple \textcolor{black}{ICPs} at the same walking speed \cite{wosnitza2013inter,mendes2013quantification,deangelis2019manifold}. 

Investigations into the existence of such a continuum in invertebrate walking are crucial for understanding the underlying control strategies used by these animals, and for any attempt to compare and contrast these strategies with those well-characterized in vertebrates. For instance, the generation of a multi-attractor system (as would be implied by the existence of discrete \textcolor{black}{stepping patterns} with discontinuous transitions between them) requires vastly different structure than that of a single-attractor system, in which prescribed \textcolor{black}{ICPs} are in fact cases along a continuum.

How is such a continuum of coordination patterns generated? Based on data gathered from both slow- and fast-walking insects (the stick insect \textit{Carausius morosus}: \cite{wendler1964laufen} and \textit{P. americana}: \cite{hughes1957co}), Wilson \cite{wilson1966insect} put forward a set of simple observations hypothesized to replicate all observed insect stepping patterns, as well as the transitions between them:
\begin{enumerate}[topsep=2pt,itemsep=-1ex,partopsep=1ex,parsep=1ex]
\item{Swing phase is initiated in a posterior to anterior wave along each ipsilateral side.}
\item{Contralateral leg pairs move in anti-phase.}
\item{The duration of swing phase within each stride is constant and independent of walking speed.}
\item{\textcolor{black}{Stride} frequency increases with speed, and is modulated by changing stance duration.}
\end{enumerate}

These `rules' \textcolor{black}{support early observations by Manton, whose extensive investigations into panarthropod walking similarly noted many common features among species \cite{manton1950peripatus,manton1952evolution,manton1952chilopoda,manton1954diplopoda,manton1972hexapod}.}  Recent work by DeAngelis et al. characterized the structure of variability in fly walking across speeds and showed that animals are able to seamlessly transition between canonical ICPs by modifying stance duration (Figure \ref{fig:transitions}), \textcolor{black}{in support of Wilson's final observation. Varying this single parameter also suffice to describe extensions of the ICP continuum beyond tripod coordination in fast-running hexapod species.} In these cases (seen, for instance in cockroaches, beetles, and ants \cite{full1991mechanics,wahl2015walking,hughes1952co}), bipod and monopod \textcolor{black}{stepping patterns} are generated via the continuously increasing overlap of the swing phases of two sets of tripod legs \cite{deangelis2019manifold}. 

Detailed behavioral studies in \textit{C. morosus} also largely agreed with Wilson's observations, and proposed a small set of locally-distributed coordination rules (`Cruse's rules') which describe how a leg affects the likelihood of the initiation of a swing event in an anterior or contralateral neighboring leg \cite{cruse1990mechanisms,durr2004behaviour}. \textit{Rule 1} states that a leg's stance-to-swing transition is suppressed while its neighbor is in swing, while \textit{Rule 2} states that the likelihood of lift-off increases once the neighboring leg touches down. While not explicitly tested in \textit{Drosophila}, it is quite likely that Cruse's rules would suffice to generate the spectrum of walking behavior characterized in flies \cite{wosnitza2013inter,mendes2013quantification}. 

Our recent work on the tardigrade \textit{H. exemplaris} confirmed the existence of these rules in the \textcolor{black}{stepping patterns of freely walking tardigrades during forward walking on agarose gel substrate \cite{nirody2021inter}.} 
\textcolor{black}{Stepping patterns in organisms with (many) more than six legs also generally follow the above observations and `rules' without undergoing the exact transitions shown in Figure \ref{fig:transitions}. For instance, similar locomotor control circuits in myriapods manifest as a metachronal wave coordination across all walking speeds, in which the phase offset between ipsilateral legs increases with increasing speed \cite{manton1950peripatus,manton1952chilopoda,manton1954diplopoda,kuroda2018dynamic,yasui2019decoding}. In these systems reducing stance duration increases the frequency of the traveling wave of swing initiations and a decrease in the number of legs involved in each cycle  $n_{\text{cycle}}$ (i.e., the `wavelength'). This results in an increase in the ipsilateral phase offset $\phi_I = \frac{1}{n_\text{cycle}}$ as walking speeds increase.} This may \textcolor{black}{further support the hypothesis} that intrinsic coordination patterns in forward walking are shared not only among insects, but across panarthropod taxa. 

\subsection*{Galloping and other such surprises}

In a set of organisms as diverse in morphology, habitat, and behavior as panarthropods, extraordinary cases will arise that deviate from any devised set of `rules'. This is inevitable regardless of how `fundamental' or general these rules purport to be. Understanding \textit{how} and \textit{why} certain examples shift away from seemingly `universal' traits often serves not only to characterize these exceptions, but to further illuminate and refine the rule. For instance, several stepping patterns observed in walking \textit{Drosophila} -- e.g., an ICP in which contralateral fore- and hind-limbs swing together while each mid-limb swings alone -- initially seemed distinct from previously-described canonical \textcolor{black}{ICPs}. However, this `non-canonical'  pattern, among several others, cleanly fits within the context of a continuum of limb coordination \cite{wilson1966insect,deangelis2019manifold}. Similarly, the same coordination rules are active in species \textcolor{black}{with more that six legs \cite{manton1950peripatus,manton1952chilopoda,manton1954diplopoda,kuroda2018dynamic,yasui2019decoding}}, as well as in insects that walk less than six legs. For instance, in case of leg loss \cite{grabowska2012quadrupedal,wosnitza2013inter} or in organisms like mantids which often hold up their forelimbs and walk with the other two pairs \cite{wilson1966insect}, the same coordination rules apply simply with the missing legs omitted. \textcolor{black}{However, the unique morphology of certain groups may drive a separation from this spectrum: the hydraulic extensor system in spider legs, for example, is believed to underlie several kinematic differences between Arachnida and other arthropod groups (for more details see, e.g., \cite{weihmann2012hydraulic,booster2015effect,hirt2017little,weihmann2020survey,boehm2021understanding}).}

\textcolor{black}{Indeed, examples that fall beyond this spectrum can be observed in several panarthropod groups. For instance, several species of arthropods (cockroaches: \cite{weihmann2017speed}, mites: \cite{weihmann2015requirements}, spiders: \cite{weihmann2013crawling}) switch to metachronal coordination at the highest observed running speeds. This switch results in a discontinuous switch in phase relationship between leg pairs and is hypothesized to be advantageous for locomotion on slippery surfaces \cite{weihmann2017speed}. In the alternating tripod, lateral ground reaction forces (GRFs) generated by the front and middle legs within each tripod brace against each other \cite{dickinson2000animals}. This may contribute to energy recovery during a stride, as well as to dynamic stability by controlling the lateral dynamics of the COM \cite{schmitt2002dynamics,weihmann2017speed}. However, these benefits are largely absent when moving on slippery or granular substrates; here, these lateral forces can risk slipping \cite{li2009sensitive}. The observed metachronal pattern, however, constitutes a desynchronization of the legs within each alternating tripod set \cite{weihmann2017speed}, removing the detrimental effects of double stance on flowing or slippery media \cite{li2009sensitive}. Furthermore, lifting the requirement for three legs to step simultaneously increases the temporal overlap between the stance periods of consecutive sets of legs, allowing for duty factor to decrease without an aerial phase \cite{weihmann2017speed}. Maintaining permanent ground contact may have additional advantages on slippery substrates, for instance because it allows for uninterrupted proprioceptive input on the animal's position with respect to the ground \cite{sponberg2008neuromechanical}.}

\textcolor{black}{One of the most prevalent features noted in ICPs across species is that swing initiations occur in a posterior to anterior wave on each ipsilateral side ($\phi_I < 0.5$); Wilson's first observation noted that this pattern manifests across all walking speeds in several insect species \cite{wilson1966insect,cruse1990mechanisms}. In fact, this `rule' holds true across panarthropods with very few exceptions, the majority of which are within the class Chilopoda (centipedes). Myriapods all progress using `locomotory waves'; millipedes (class Diplopoda) and two of the five orders of centipedes display the expected posterior-to-anterior pattern \cite{manton1954diplopoda,kuroda2018dynamic}. However, the other three centipede orders (Craterostigmorpha, Scolopendromorpha, Geophilomorpha) exhibit \textit{retrograde} waves: swing initiations that occur in an anterior-to-posterior manner ($\phi_I > 0.5$) \cite{manton1952chilopoda,kuroda2018dynamic}. Molecular phylogenies of Myriapoda indicate that retrograde waves may be a derived feature \cite{miyazawa2014molecular,fernandez2018phylogenomics}; further ecological, functional, and anatomical studies into these centipede orders will be needed to understand both the selective factors and the underlying neural basis for the determination of wave direction.}

\textcolor{black}{Contralateral coupling is generally more variable than coupling between ipsilateral leg pairs, both within a single species and among different panarthropod species. In particular, several species across diverse panarthropod taxa exhibit in-phase contralateral coordination, rather than anti-phase as in Wilson's second observation \cite{wilson1966insect}.} Observations in three species of flightless dung beetles in the genus \textit{Pachysoma} noted a `galloping' \textcolor{black}{coordination pattern} in which contralateral leg pairs step in-phase with each other \cite{smolka2013new}. Previously, synchronous contralateral coordination in terrestrial arthropods has only been observed transiently, e.g., when traversing three-dimensional terrain or for the first few steps when walking is first initiated  \cite{pearson1984characteristics}. Aquatic species (e.g., krill \cite{zhang2014neural} \textcolor{black}{and water striders \cite{bowdan1978walking}}) display in-phase contralateral strokes \textcolor{black}{while moving under or on the surface of water}, a coordination pattern believed to be highly optimized for aquatic locomotion \cite{zhang2014neural,takagi2015swimming}. \textcolor{black}{Many species of millipedes similarly show a preference for in-phase contralateral coupling, a pattern which has been measured to provide increased pushing force during burrowing \cite{manton1954diplopoda,manton1958diplopoda}.} Interestingly, galloping species of \textit{Pachysoma} are not faster than their tripodal siblings, suggesting that there is no speed advantage to this \textcolor{black}{stepping pattern} \cite{smolka2013new}. In support of the hypothesis that in-phase contralateral swings may provide some advantage on shifting substrates like the sands desert-dwelling \textit{Pachysoma} must traverse, we observed sustained `gallops' in tardigrades walking on substrates of reduced stiffness ($\sim$10kPa) \textcolor{black}{\cite{nirody2021inter}}. 

\subsection*{A simple framework for the panarthropod ICP continuum} 

\begin{figure}[t!]
\begin{center}
\includegraphics[width=0.95\textwidth]{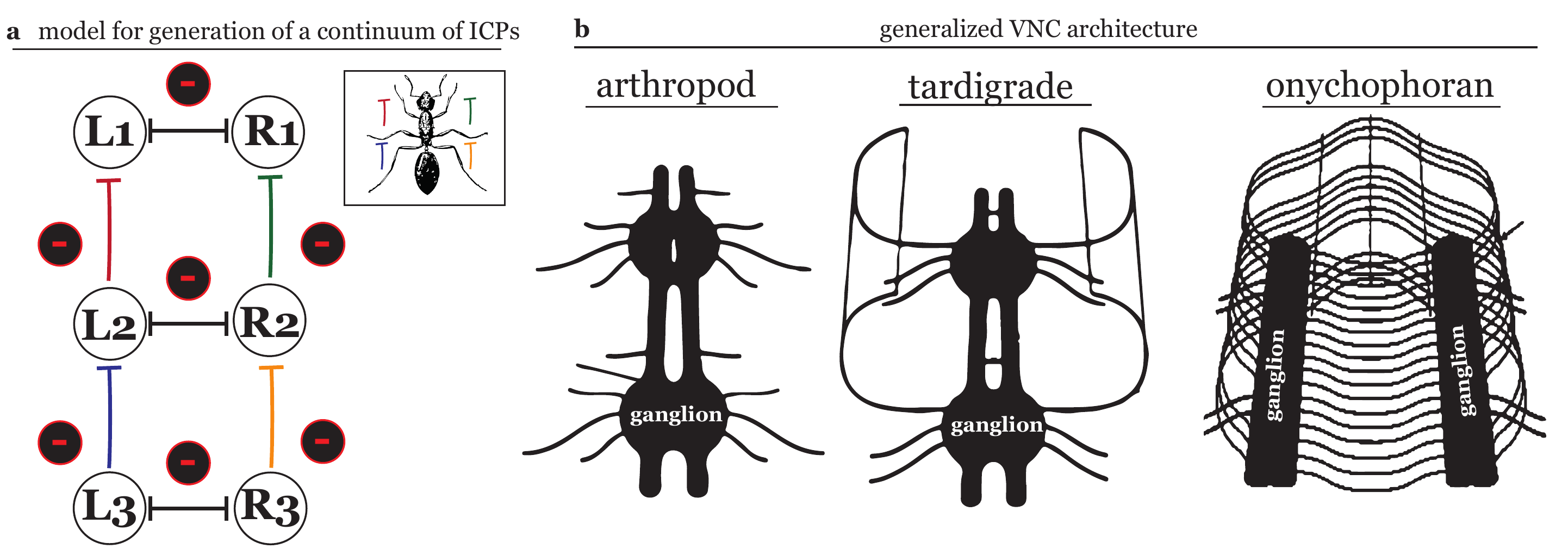}
\caption{\textit{A simple model for the generation of inter-leg coordination patterns (ICPs) based on ventral nerve cord (VNC) architecture.} \textbf{(a)} Detailed characterization of \textit{Drosophila} coordination patterns across walking speeds in \cite{deangelis2019manifold} suggests that a simple control circuit may be sufficient to generate all observed ICPs in walking \textit{Drosophila}. The proposed circuit comprises mutual inhibitory connections between contralateral leg pairs and a posterior-to-anterior inhibition on each ipsilateral side; \textcolor{black}{inhibitory connections are denoted by capped vertical lines with associated $(-)$ signs}. These connections are postulated to be found in the thoracic ganglia of the \textit{Drosophila} VNC. \textbf{(b)} VNC structure in \textcolor{black}{arthropods and tardigrades} consists of a series of segmented ganglia, each of which corresponds to a single leg pair. \textcolor{black}{Onychophorans have two laterally-located ganglia connected by median commissures at each leg pair.} The conserved general topology of VNC architecture across Panarthropoda lends support to the possibility that the functional similarities in \textcolor{black}{stepping} patterns observed in these diverse taxa may originate from a shared underlying control strategy. \textbf{(b)} modified from~\cite{yang2016fuxianhuiid}.}
\label{fig:vnccomp}
\end{center}
\end{figure}

A large variety of theoretical and computational models have been developed over the years to describe stepping patterns in hexapod locomotion \cite{cruse1990mechanisms,ijspeert2008central,aminzare2018gait,szczecinski2018static,schilling2020decentralized}. Based on a comprehensive analysis of the variability in \textit{Drosophila} leg coordination across walking speeds, DeAngelis et al.~propose that a single continuum can describe all observed patterns in fly walking \cite{deangelis2019manifold}. As previously mentioned, such a continuum, which does not need to account for multiple discrete coordination patterns, allows for the possibility of a simpler control circuit underlying forward walking in \textit{Drosophila}. This simple model suggests the existence of mutual inhibitory coupling between contralateral neuropil and posterior-to-anterior inhibitory coupling between ipsilateral neuropil of the ventral nerve cord (VNC) in \textit{Drosophila} (Figure \ref{fig:vnccomp}a). Excitingly, DeAngelis et al.~also show that varying a single parameter, stance duration, can replicate fundamental components of the observed spectrum of ICPs without any speed-dependent modulation of ipsilateral and contralateral coupling~\textcolor{black}{(Figure \ref{fig:transitions})}.

The inter-leg coordination patterns observed in the \textit{Drosophila} continuum closely mirror features of those in a range of insects and other arthropods, as well as those recently characterized in tardigrades (Figure \ref{fig:coord}a). 
\textcolor{black}{Panarthropod groups display notable similarities in VNC architecture (Figure \ref{fig:vnccomp}b). This may intriguingly support the existence of a shared underlying locomotor control circuit in Panarthropoda, which has been modified along certain clades due to specific pressures on organismal performance \cite{yang2016fuxianhuiid}. The VNC in onychophorans shows several differences from that of tardigrades and arthropods, containing ladder-like lateral ganglia connected by interpedal median commissures (Figure \ref{fig:vnccomp}b). However, the topology of this structure is not significantly different from the segmented hemiganglia of tardigrades and arthropods and does not rule out the existence of a shared control circuit between onychophorans and the other panarthropod taxa \cite{yang2016fuxianhuiid}. Previous observations of onycophoran locomotion determined that average ipsilateral phase offsets are consistent with those of other panarthropods (Figure~\ref{fig:coord}a); coupling between contralateral leg pairs is irregular at low speeds but converges to anti-phase contralateral coupling at high speeds \cite{manton1950peripatus,manton1952evolution,oliveira2019functional}. More detailed analyses in velvet worm species are needed to reveal how, if at all, morphological differences between the VNC of Onychophora and Tardigrada+Arthropoda affect inter-leg coordination.} 

\textcolor{black}{Molecular studies have found that the compact tardigrade body plan evolved from a loss of a large body region corresponding to the entire thorax and part of the abdomen in arthropods. This indicates that the tardigrades' legged segments are homologous only with the head region of other panarthropods \cite{smith2016compact}. These results support the hypothesis that the diversity of head appendages in arthropods and onychophorans evolved from legs \cite{eriksson2010head,angelini2012extent}. While this does not necessarily preclude the idea that a common circuit underlies forward walking in panarthropods, the alternative hypothesis is that similarities in tardigrade and arthropod coordination patterns have independently evolved. This parallel convergence onto similar inter-leg coordination strategies by these two groups is intriguing given their remarkable disparities in size and skeletal structure, and may provide significant insight into general design principles for efficient and robust control of multi-legged locomotion. A more definitive distinction between these scenarios will require deeper functional studies combined with molecular and phylogenetic analyses.}

In accordance with observations made by Wilson decades prior \cite{wilson1966insect}, we note several key features of an `idealized' ICP spectrum. First, as noted previously, only stance duration is varied with walking speed; the duration of the swing phase is speed-independent. Second, ipsilateral swings do not overlap and occur in a posterior-to-anterior wave. \textcolor{black}{In an animal with $n$ legs, this results in the phase offset between ipsilateral legs increasing from $\phi_I \geq \frac{1}{n}$ at the lowest walking speeds up until a maximum offset of $\phi_I = 0.5$ at the fastest speeds ($\phi_I > 0.5$ corresponds to a retrograde wave of swing initiations that travels posteriorly). In the case of hexapods, this corresponds to a speed-dependent continuum} varying smoothly from $\phi_I = \frac{1}{6}$ in wave \textcolor{black}{coordination} to $\phi_I = \frac{1}{3}$ in tetrapod to $\phi_I = \frac{1}{2}$ in tripod coordination \textcolor{black}{(Figure \ref{fig:transitions})}. Finally, contralateral leg pairs show a preference for anti-phase coordination $\phi_C = \frac{1}{2}$ across all walking speeds (Figure \ref{fig:coord}b).

Of course, measurements in freely behaving animals rarely adhere to any semblance of `ideal' relationships. One such deviation arises from the stipulation that contralateral legs prefer an anti-phase coordination. The `canonical' tetrapod comprises a sequence of swing initiations by three sets of two legs; this results in a contralateral offset of $\phi_C = \frac{1}{3}$ (or $\phi_C = \frac{2}{3}$ for the mirror-image tetrapod) at lower speeds. DeAngelis et al. \cite{deangelis2019manifold} report a cross-body offset in step timing such that limbs that are meant to swing simultaneously are actually slightly offset in time, resulting in an anti-phase contralateral phase relationship \textcolor{black}{(Figure \ref{fig:transitions})}. Nearly all surveyed arthropod species similarly showed, on average, anti-phase contralateral phasing \textcolor{black}{(Figure \ref{fig:coord}b; though several exceptions are noted in the section above)}. Of course, this may result from a bimodal distribution with peaks at $\frac{1}{3}$ and $\frac{2}{3}$ corresponding to the two mirror-image tetrapods. However, all studies in which complete data was made available reported contralateral phase-offset distributions with a single peak centered around $\phi_C = \frac{1}{2}$. Further investigation into this relationship across taxa will be needed to confirm the generalizability of this simple model. 

Measured inter-leg relationships also show high variability (Figure \ref{fig:coord}). Interestingly, all pairwise inter-leg relationships show higher variability at low speeds than during fast walking. A possible explanation for this pattern  is purely physical: when limbs have asymmetric duty cycles, they cover different fractions of a cycle per unit time when in swing vs stance. As such, slow walking, which has longer stance phases and approximately the same swing duration as fast walking, will show greater variance in relative phasing \cite{couzin2015intersegmental,deangelis2019manifold}. 

An alternative is that inter-limb coupling is more affected by sensory information at low speeds than at high speeds, and thus is more variable \cite{schilling2020decentralized}. This explanation is consistent with observations of higher variability in limb coordination patterns in slow-walking insects when compared to fast runners \cite{delcomyn1991perturbation,sponberg2008neuromechanical,bidaye2018six}. We note that this option does not necessarily require a speed-dependent modulation of inter-limb coupling strength: there is a fundamental timescale related to the propagation of sensory information, which is too slow to drive behavior at speeds higher than approximately 5 \textcolor{black}{strides per second} (corresponding to a stepping period of approximately 200 ms) in \textit{P. americana} \cite{delcomyn1991perturbation,schilling2020decentralized}; this limit may be higher in \textit{Drosophila} due to its relatively smaller size. 

Contralateral coordination is generally weaker than ipsilateral coordination across all surveyed species \textcolor{black}{\cite{nirody2021inter}}. Studies in a range of organisms have shown flexibility in the coupling between contralateral leg pairs within single individuals in response to external stimuli; for instance, we characterized a transition from anti-phase to in-phase contralateral coupling in \textit{H. exemplaris} with changes in substrate stiffness with no shift observed in ipsilateral phase offsets \textcolor{black}{\cite{nirody2021inter}}. \textcolor{black}{Furthermore, the relative weakness of coupling between contralateral leg pairs in comparison to ipsilateral leg pair coupling is consistent with the hypothesis that the underlying controller proposed based on \textit{Drosophila} is shared across panarthropods. Contralateral phasing is quite variable across taxa, ranging from in-phase in swimming Crustacea \cite{zhang2014neural} to anti-phase in running insects \cite{full1991mechanics,merienne2021dynamics} and arachnids \cite{spagna2011gait,weihmann2015requirements}. However, ipsilateral phase relationships are consistent across nearly all characterized species (with few exceptions; see, e.g, \cite{manton1952chilopoda,kuroda2018dynamic}).}

In order to further substantiate how the walking system characterized in \textit{Drosophila} compares to that in other insects, and for panarthropods in general, it will be necessary to undertake deeper comparative investigations. We note that this review focused on \textcolor{black}{leg kinematics during} forward walking on flat surfaces; far less comparative data was available for \textcolor{black}{body and center-of-mass dynamics, as well as for} more complex behavior such as turning, obstacle traversal, backward walking, and loaded locomotion. Intriguingly, there is evidence that turning in \textit{Drosophila} requires only a small modification of the hypothesized forward walking circuit~\cite{deangelis2019manifold}; this remains to be tested in other animals. As tools for automating collection and analysis of large behavioral datasets become more commonplace~\cite{mendes2013quantification,pereira2019fast,deangelis2019manifold}, the goal of intensive and comprehensive characterization of walking across panarthropod taxa comes within reach. However, crucial to the success of such studies is the accessibility of raw movement data in a wide range of species; it is our hope that this work sheds light on the importance of these analyses.

\bibliographystyle{plain}
\bibliography{references.bib}

\begin{thebibliography}{10}

\bibitem{alexander1989optimization}
RM~Alexander.
\newblock Optimization and gaits in the locomotion of vertebrates.
\newblock {\em Physiological Reviews}, 69(4):1199--1227, 1989.

\bibitem{alexander1983dynamic}
RM~Alexander and AS~Jayes.
\newblock A dynamic similarity hypothesis for the gaits of quadrupedal mammals.
\newblock {\em Journal of Zoology}, 201(1):135--152, 1983.

\bibitem{aminzare2018gait}
Zahra Aminzare, Vaibhav Srivastava, and Philip Holmes.
\newblock Gait transitions in a phase oscillator model of an insect central
  pattern generator.
\newblock {\em SIAM Journal on Applied Dynamical Systems}, 17(1):626--671,
  2018.

\bibitem{angelini2012extent}
David~R Angelini, Frank~W Smith, and Elizabeth~L Jockusch.
\newblock Extent with modification: leg patterning in the beetle tribolium
  castaneum and the evolution of serial homologs.
\newblock {\em G3: Genes| Genomes| Genetics}, 2(2):235--248, 2012.

\bibitem{ayali2015comparative}
Amir Ayali, Anke Borgmann, Ansgar B{\"u}schges, Einat Couzin-Fuchs, Silvia
  Daun-Gruhn, and Philip Holmes.
\newblock The comparative investigation of the stick insect and cockroach
  models in the study of insect locomotion.
\newblock {\em Current Opinion in Insect Science}, 12:1--10, 2015.

\bibitem{bender2011kinematic}
John~A Bender, Elaine~M Simpson, Brian~R Tietz, Kathryn~A Daltorio, Roger~D
  Quinn, and Roy~E Ritzmann.
\newblock Kinematic and behavioral evidence for a distinction between trotting
  and ambling gaits in the cockroach blaberus discoidalis.
\newblock {\em Journal of Experimental Biology}, 214(12):2057--2064, 2011.

\bibitem{bidaye2018six}
Salil~S Bidaye, Till Bockem{\"u}hl, and Ansgar B{\"u}schges.
\newblock Six-legged walking in insects: how cpgs, peripheral feedback, and
  descending signals generate coordinated and adaptive motor rhythms.
\newblock {\em Journal of neurophysiology}, 119(2):459--475, 2018.

\bibitem{boehm2021understanding}
Charlotte Boehm, Johanna Schultz, and Christofer Clemente.
\newblock Understanding the limits to the hydraulic leg mechanism: the effects
  of speed and size on limb kinematics in vagrant arachnids.
\newblock {\em Journal of Comparative Physiology A}, pages 1--12, 2021.

\bibitem{booster2015effect}
NA~Booster, FY~Su, SC~Adolph, and AN~Ahn.
\newblock Effect of temperature on leg kinematics in sprinting tarantulas
  (aphonopelma hentzi): high speed may limit hydraulic joint actuation.
\newblock {\em Journal of Experimental Biology}, 218(7):977--982, 2015.

\bibitem{bowdan1978walking}
Elizabeth Bowdan.
\newblock Walking and rowing in the water strider, gerris remigis.
\newblock {\em Journal of comparative physiology}, 123(1):51--57, 1978.

\bibitem{clifton2020uneven}
GT~Clifton, D~Holway, and N~Gravish.
\newblock Uneven substrates constrain walking speed in ants through modulation
  of stride frequency more than stride length.
\newblock {\em Royal Society open science}, 7(3):192068, 2020.

\bibitem{couzin2015intersegmental}
Einat Couzin-Fuchs, Tim Kiemel, Omer Gal, Amir Ayali, and Philip Holmes.
\newblock Intersegmental coupling and recovery from perturbations in freely
  running cockroaches.
\newblock {\em Journal of Experimental Biology}, 218(2):285--297, 2015.

\bibitem{cruse1990mechanisms}
Holk Cruse.
\newblock What mechanisms coordinate leg movement in walking arthropods?
\newblock {\em Trends in Neurosciences}, 13(1):15--21, 1990.

\bibitem{dallmann2019motor}
Chris~J Dallmann, Volker D{\"u}rr, and Josef Schmitz.
\newblock Motor control of an insect leg during level and incline walking.
\newblock {\em Journal of Experimental Biology}, 222(7), 2019.

\bibitem{dallmann2017load}
Chris~J Dallmann, Thierry Hoinville, Volker D{\"u}rr, and Josef Schmitz.
\newblock A load-based mechanism for inter-leg coordination in insects.
\newblock {\em Proceedings of the Royal Society B: Biological Sciences},
  284(1868):20171755, 2017.

\bibitem{deangelis2019manifold}
Brian~D DeAngelis, Jacob~A Zavatone-Veth, and Damon~A Clark.
\newblock The manifold structure of limb coordination in walking drosophila.
\newblock {\em eLife}, 8:e46409, 2019.

\bibitem{delcomyn1991perturbation}
Fred Delcomyn.
\newblock Perturbation of the motor system in freely walking cockroaches. i.
  rear leg amputation and the timing of motor activity in leg muscles.
\newblock {\em Journal of Experimental Biology}, 156(1):483--502, 1991.

\bibitem{dickinson2000animals}
Michael~H Dickinson, Claire~T Farley, Robert~J Full, MAR Koehl, Rodger Kram,
  and Steven Lehman.
\newblock How animals move: an integrative view.
\newblock {\em Science}, 288(5463):100--106, 2000.

\bibitem{durr2004behaviour}
Volker D{\"u}rr, Josef Schmitz, and Holk Cruse.
\newblock Behaviour-based modelling of hexapod locomotion: linking biology and
  technical application.
\newblock {\em Arthropod Structure \& Development}, 33(3):237--250, 2004.

\bibitem{durr2018motor}
Volker D{\"u}rr, Leslie~M Theunissen, Chris~J Dallmann, Thierry Hoinville, and
  Josef Schmitz.
\newblock Motor flexibility in insects: adaptive coordination of limbs in
  locomotion and near-range exploration.
\newblock {\em Behavioral ecology and sociobiology}, 72(1):1--21, 2018.

\bibitem{eriksson2010head}
Bo~Joakim Eriksson, Noel~N Tait, Graham~E Budd, Ralf Janssen, and Michael Akam.
\newblock Head patterning and hox gene expression in an onychophoran and its
  implications for the arthropod head problem.
\newblock {\em Development genes and evolution}, 220(3):117--122, 2010.

\bibitem{fernandez2018phylogenomics}
Rosa Fern{\'a}ndez, Gregory~D Edgecombe, and Gonzalo Giribet.
\newblock Phylogenomics illuminates the backbone of the myriapoda tree of life
  and reconciles morphological and molecular phylogenies.
\newblock {\em Scientific reports}, 8(1):1--7, 2018.

\bibitem{full1990mechanics}
Robert~J Full and Michael~S Tu.
\newblock Mechanics of six-legged runners.
\newblock {\em Journal of experimental biology}, 148(1):129--146, 1990.

\bibitem{full1991mechanics}
Robert~J Full and Michael~S Tu.
\newblock Mechanics of a rapid running insect: two-, four-and six-legged
  locomotion.
\newblock {\em Journal of Experimental Biology}, 156(1):215--231, 1991.

\bibitem{goldman2006dynamics}
Daniel~I Goldman, Tao~S Chen, Daniel~M Dudek, and Robert~J Full.
\newblock Dynamics of rapid vertical climbing in cockroaches reveals a
  template.
\newblock {\em Journal of Experimental Biology}, 209(15):2990--3000, 2006.

\bibitem{grabowska2012quadrupedal}
Martyna Grabowska, Elzbieta Godlewska, Joachim Schmidt, and Silvia Daun-Gruhn.
\newblock Quadrupedal gaits in hexapod animals--inter-leg coordination in
  free-walking adult stick insects.
\newblock {\em Journal of Experimental Biology}, 215(24):4255--4266, 2012.

\bibitem{graham1972behavioural}
D~Graham.
\newblock A behavioural analysis of the temporal organisation of walking
  movements in the 1st instar and adult stick insect (carausius morosus).
\newblock {\em Journal of Comparative Physiology}, 81(1):23--52, 1972.

\bibitem{heglund1988speed}
Norman~C Heglund and C~Richard Taylor.
\newblock Speed, stride frequency and energy cost per stride: how do they
  change with body size and gait?
\newblock {\em Journal of Experimental Biology}, 138(1):301--318, 1988.

\bibitem{heglund1974scaling}
Norman~C Heglund, C~Richard Taylor, and Thomas~A McMahon.
\newblock Scaling stride frequency and gait to animal size: mice to horses.
\newblock {\em Science}, 186(4169):1112--1113, 1974.

\bibitem{hirt2017little}
Myriam~R Hirt, Tobias Lauermann, Ulrich Brose, Lucas~PJJ Noldus, and Anthony~I
  Dell.
\newblock The little things that run: a general scaling of invertebrate
  exploratory speed with body mass.
\newblock {\em Ecology}, 2017.

\bibitem{hoyt1981gait}
Donald~F Hoyt and C~Richard Taylor.
\newblock Gait and the energetics of locomotion in horses.
\newblock {\em Nature}, 292(5820):239--240, 1981.

\bibitem{hoyt2006relations}
Donald~F Hoyt, Steven~J Wickler, Darren~J Dutto, Gwenn~E Catterfeld, and Devin
  Johnsen.
\newblock What are the relations between mechanics, gait parameters, and
  energetics in terrestrial locomotion?
\newblock {\em Journal of Experimental Zoology Part A: Comparative Experimental
  Biology}, 305(11):912--922, 2006.

\bibitem{hughes1952co}
George~M Hughes.
\newblock The co-ordination of insect movements: I the walking movements of
  insects.
\newblock {\em Journal of Experimental Biology}, 29(2):267--285, 1952.

\bibitem{hughes1957co}
GM~Hughes.
\newblock The co-ordination of insect movements: 11. the effect of limb
  amputation and the cutting of commissures in the cockroach (blatta
  oiuentalis).
\newblock {\em Journal of Experimental Biology}, 34(3):306--333, 1957.

\bibitem{ijspeert2008central}
Auke~Jan Ijspeert.
\newblock Central pattern generators for locomotion control in animals and
  robots: a review.
\newblock {\em Neural networks}, 21(4):642--653, 2008.

\bibitem{kuroda2018dynamic}
Shigeru Kuroda, Nariya Uchida, and Toshiyuki Nakagaki.
\newblock Dynamic gait transition in the scolopendromorpha scolopocryptops
  rubiginosus l. koch centipede.
\newblock {\em bioRxiv}, page 312280, 2018.

\bibitem{li2009sensitive}
Chen Li, Paul~B Umbanhowar, Haldun Komsuoglu, Daniel~E Koditschek, and Daniel~I
  Goldman.
\newblock Sensitive dependence of the motion of a legged robot on granular
  media.
\newblock {\em Proceedings of the National Academy of Sciences},
  106(9):3029--3034, 2009.

\bibitem{manton1950peripatus}
SM~Manton.
\newblock The evolution of arthropodan locomotory mechanisms.—part 1. the
  locomotion of {P}eripatus.
\newblock {\em Zoological Journal of the Linnean Society}, 41(282):529--570,
  1950.

\bibitem{manton1952evolution}
SM~Manton.
\newblock The evolution of arthropodan locomotory mechanisms.—part 2. general
  introduction to the locomotory mechanisms of the {A}rthropoda.
\newblock {\em Zoological journal of the Linnean Society}, 42(284):93--117,
  1952.

\bibitem{manton1952chilopoda}
SM~Manton.
\newblock The evolution of arthropodan locomotory mechanisms—part 3. the
  locomotion of the {C}hilopoda and {P}auropoda.
\newblock {\em Zoological Journal of the Linnean Society}, 42(284):118--167,
  1952.

\bibitem{manton1954diplopoda}
SM~Manton.
\newblock The evolution of arthropodan locomotory mechanisms.—part 4. the
  structure, habits and evolution of the {D}iplopoda.
\newblock {\em Zoological Journal of the Linnean Society}, 42(286):299--368,
  1954.

\bibitem{manton1958diplopoda}
SM~Manton.
\newblock The evolution of arthropodan locomotory mechanisms. part 6. habits
  and evolution of the lysiopetaloidea [{D}iplopoda], some principles of leg
  design in {D}iplopoda and {C}hilopoda, and limb structure of {D}iplopoda.
\newblock {\em Zoological Journal of the Linnean Society}, 43(293):487--556,
  1958.

\bibitem{manton1972hexapod}
SM~Manton.
\newblock The evolution of arthropodan locomotory mechanisms: Part 10.
  locomotory habits, morphology and evolution of the hexapod classes.
\newblock {\em Zoological Journal of the Linnean Society}, 51(3-4):203--400,
  1972.

\bibitem{mendes2013quantification}
C{\'e}sar~S Mendes, Imre Bartos, Turgay Akay, Szabolcs M{\'a}rka, and Richard~S
  Mann.
\newblock Quantification of gait parameters in freely walking wild type and
  sensory deprived drosophila melanogaster.
\newblock {\em eLife}, 2:e00231, 2013.

\bibitem{merienne2021dynamics}
Hugo Merienne, G{\'e}rard Latil, Pierre Moretto, and Vincent Fourcassi{\'e}.
\newblock Dynamics of locomotion in the seed harvesting ant messor barbarus:
  effect of individual body mass and transported load mass.
\newblock {\em PeerJ}, 9:e10664, 2021.

\bibitem{miyazawa2014molecular}
Hideyuki Miyazawa, Chiaki Ueda, Kensuke Yahata, and Zhi-Hui Su.
\newblock Molecular phylogeny of myriapoda provides insights into evolutionary
  patterns of the mode in post-embryonic development.
\newblock {\em Scientific reports}, 4(1):1--9, 2014.

\bibitem{nirody2021inter}
Jasmine~A Nirody, Lisset~A Duran, Deborah Johnston, and Daniel~J Cohen.
\newblock Tardigrades exhibit robust inter-limb coordination across walking
  speeds.
\newblock {\em bioRxiv}, 2021.

\bibitem{nishii2000legged}
Jun Nishii.
\newblock Legged insects select the optimal locomotor pattern based on the
  energetic cost.
\newblock {\em Biological cybernetics}, 83(5):435--442, 2000.

\bibitem{niven2008diversity}
Jeremy~E Niven, Christopher~M Graham, and Malcolm Burrows.
\newblock Diversity and evolution of the insect ventral nerve cord.
\newblock {\em Annu. Rev. Entomol.}, 53:253--271, 2008.

\bibitem{oliveira2019functional}
Ivo de~Sena Oliveira, Andreas Kumerics, Henry Jahn, Mark M{\"u}ller, Franz
  Pfeiffer, and Georg Mayer.
\newblock Functional morphology of a lobopod: case study of an onychophoran
  leg.
\newblock {\em Royal Society Open Science}, 6(10):191200.

\bibitem{pearson1984characteristics}
KG~Pearson and R~Franklin.
\newblock Characteristics of leg movements and patterns of coordination in
  locusts walking on rough terrain.
\newblock {\em The International Journal of Robotics Research}, 3(2):101--112,
  1984.

\bibitem{pearson1973nervous}
KG~Pearson and JF~Iles.
\newblock Nervous mechanisms underlying intersegmental co-ordination of leg
  movements during walking in the cockroach.
\newblock {\em Journal of Experimental Biology}, 58(3):725--744, 1973.

\bibitem{pereira2019fast}
Talmo~D Pereira, Diego~E Aldarondo, Lindsay Willmore, Mikhail Kislin, Samuel
  S-H Wang, Mala Murthy, and Joshua~W Shaevitz.
\newblock Fast animal pose estimation using deep neural networks.
\newblock {\em Nature Methods}, 16(1):117--125, 2019.

\bibitem{pfeffer2019high}
Sarah~Elisabeth Pfeffer, Verena~Luisa Wahl, Matthias Wittlinger, and Harald
  Wolf.
\newblock High-speed locomotion in the saharan silver ant, cataglyphis
  bombycina.
\newblock {\em Journal of Experimental Biology}, 222(20), 2019.

\bibitem{digitize}
T.~Poisot.
\newblock The digitize package: extracting numerical data from scatterplots.
\newblock {\em The R Journal}, 3(1):25--26, 2011.

\bibitem{reinhardt2014level}
Lars Reinhardt and Reinhard Blickhan.
\newblock Level locomotion in wood ants: evidence for grounded running.
\newblock {\em Journal of Experimental Biology}, 217(13):2358--2370, 2014.

\bibitem{schilling2020decentralized}
Malte Schilling and Holk Cruse.
\newblock Decentralized control of insect walking: A simple neural network
  explains a wide range of behavioral and neurophysiological results.
\newblock {\em PLoS Computational Biology}, 16(4):e1007804, 2020.

\bibitem{schmitt2002dynamics}
John Schmitt, Mariano Garcia, RC~Razo, Philip Holmes, and Robert~J Full.
\newblock Dynamics and stability of legged locomotion in the horizontal plane:
  a test case using insects.
\newblock {\em Biological cybernetics}, 86(5):343--353, 2002.

\bibitem{smarandache2016arthropod}
Carmen~Ramona Smarandache-Wellmann.
\newblock Arthropod neurons and nervous system.
\newblock {\em Current Biology}, 26(20):R960--R965, 2016.

\bibitem{smith2016compact}
Frank~W Smith, Thomas~C Boothby, Ilaria Giovannini, Lorena Rebecchi,
  Elizabeth~L Jockusch, and Bob Goldstein.
\newblock The compact body plan of tardigrades evolved by the loss of a large
  body region.
\newblock {\em Current Biology}, 26(2):224--229, 2016.

\bibitem{smolka2013new}
Jochen Smolka, Marcus~J Byrne, Clarke~H Scholtz, and Marie Dacke.
\newblock A new galloping gait in an insect.
\newblock {\em Current Biology}, 23(20):R913--R915, 2013.

\bibitem{spagna2011gait}
Joseph~C Spagna, Edgar~A Valdivia, and Vivin Mohan.
\newblock Gait characteristics of two fast-running spider species (hololena
  adnexa and hololena curta), including an aerial phase (araneae: Agelenidae).
\newblock {\em The Journal of Arachnology}, 39(1):84--91, 2011.

\bibitem{sponberg2008neuromechanical}
S~Sponberg and RJ~Full.
\newblock Neuromechanical response of musculo-skeletal structures in
  cockroaches during rapid running on rough terrain.
\newblock {\em Journal of Experimental Biology}, 211(3):433--446, 2008.

\bibitem{starke2009walk}
Sandra~D Starke, Justine~J Robilliard, Renate Weller, Alan~M Wilson, and Thilo
  Pfau.
\newblock Walk--run classification of symmetrical gaits in the horse: a
  multidimensional approach.
\newblock {\em Journal of the Royal Society Interface}, 6(33):335--342, 2009.

\bibitem{storch2009crustacea}
Volker Storch and Ulrich Welsch.
\newblock Crustacea, krebse.
\newblock {\em K{\"u}kenthal—Zoologisches Praktikum}, pages 213--244, 2009.

\bibitem{szczecinski2018static}
Nicholas~S Szczecinski, Till Bockem{\"u}hl, Alexander~S Chockley, and Ansgar
  B{\"u}schges.
\newblock Static stability predicts the continuum of interleg coordination
  patterns in drosophila.
\newblock {\em Journal of Experimental Biology}, 221(22), 2018.

\bibitem{takagi2015swimming}
Daisuke Takagi.
\newblock Swimming with stiff legs at low reynolds number.
\newblock {\em Physical Review E}, 92(2):023020, 2015.

\bibitem{ting1994dynamic}
LH~Ting, Reinhard Blickhan, and Robert~J Full.
\newblock Dynamic and static stability in hexapedal runners.
\newblock {\em Journal of Experimental Biology}, 197(1):251--269, 1994.

\bibitem{wahl2015walking}
Verena Wahl, Sarah~E Pfeffer, and Matthias Wittlinger.
\newblock Walking and running in the desert ant cataglyphis fortis.
\newblock {\em Journal of Comparative Physiology A}, 201(6):645--656, 2015.

\bibitem{weihmann2013crawling}
Tom Weihmann.
\newblock Crawling at high speeds: steady level locomotion in the spider
  cupiennius salei—global kinematics and implications for centre of mass
  dynamics.
\newblock {\em PLoS One}, 8(6):e65788, 2013.

\bibitem{weihmann2020survey}
Tom Weihmann.
\newblock Survey of biomechanical aspects of arthropod
  terrestrialisation--substrate bound legged locomotion.
\newblock {\em Arthropod Structure \& Development}, 59:100983, 2020.

\bibitem{weihmann2017speed}
Tom Weihmann, Pierre-Guillaume Brun, and Emily Pycroft.
\newblock Speed dependent phase shifts and gait changes in cockroaches running
  on substrates of different slipperiness.
\newblock {\em Frontiers in Zoology}, 14(1):1--15, 2017.

\bibitem{weihmann2015requirements}
Tom Weihmann, Hanns~Hagen Goetzke, and Michael G{\"u}nther.
\newblock Requirements and limits of anatomy-based predictions of locomotion in
  terrestrial arthropods with emphasis on arachnids.
\newblock {\em Journal of Paleontology}, 89(6):980--990, 2015.

\bibitem{weihmann2012hydraulic}
Tom Weihmann, Michael G{\"u}nther, and Reinhard Blickhan.
\newblock Hydraulic leg extension is not necessarily the main drive in large
  spiders.
\newblock {\em Journal of Experimental Biology}, 215(4):578--583, 2012.

\bibitem{wendler1964laufen}
Gernot Wendler.
\newblock Laufen und stehen der stabheuschrecke carausius morosus:
  sinnesborstenfelder in den beingelenken als glieder von regelkreisen.
\newblock {\em Zeitschrift f{\"u}r vergleichende Physiologie}, 48(2):198--250,
  1964.

\bibitem{wilson1966insect}
Donald~M Wilson.
\newblock Insect walking.
\newblock {\em Annual review of entomology}, 11(1):103--122, 1966.

\bibitem{wosnitza2013inter}
Anne Wosnitza, Till Bockem{\"u}hl, Michael D{\"u}bbert, Henrike Scholz, and
  Ansgar B{\"u}schges.
\newblock Inter-leg coordination in the control of walking speed in drosophila.
\newblock {\em Journal of Experimental Biology}, 216(3):480--491, 2013.

\bibitem{yang2016fuxianhuiid}
Jie Yang, Javier Ortega-Hern{\'a}ndez, Nicholas~J Butterfield, Yu~Liu, George~S
  Boyan, Jin-bo Hou, Tian Lan, and Xi-guang Zhang.
\newblock Fuxianhuiid ventral nerve cord and early nervous system evolution in
  panarthropoda.
\newblock {\em Proceedings of the National Academy of Sciences},
  113(11):2988--2993, 2016.

\bibitem{yasui2019decoding}
Kotaro Yasui, Takeshi Kano, Emily~M Standen, Hitoshi Aonuma, Auke~J Ijspeert,
  and Akio Ishiguro.
\newblock Decoding the essential interplay between central and peripheral
  control in adaptive locomotion of amphibious centipedes.
\newblock {\em Scientific reports}, 9(1):1--11, 2019.

\bibitem{zhang2014neural}
Calvin Zhang, Robert~D Guy, Brian Mulloney, Qinghai Zhang, and Timothy~J Lewis.
\newblock Neural mechanism of optimal limb coordination in crustacean swimming.
\newblock {\em Proceedings of the National Academy of Sciences},
  111(38):13840--13845, 2014.

\end{thebibliography}

\end{document}